\documentclass[preprint,11pt]{elsarticle}
\usepackage{graphicx}

\usepackage{amssymb}

\usepackage{amsthm}

\journal{Nuclear Physics B}
\newcommand{\ba}{\begin{eqnarray}}
\newcommand{\ea}{\end{eqnarray}}
\newcommand{\ban}{\begin{eqnarray*}}
\newcommand{\ean}{\end{eqnarray*}}
\newcommand{\be}{\begin{equation}}
\newcommand{\ee}{\end{equation}}
\begin{document}

\begin{frontmatter}
\title{CP violation effects on the neutrino degeneracy parameters in the Early Universe}
\author{J\'er\^ome Gava and Cristina Volpe}
\ead{gava@ipno.in2p3.fr,volpe@ipno.in2p3.fr}
\address[label1]{Institut de Physique Nucl\'eaire, F-91406 Orsay cedex, 
CNRS/IN2P3 and University of Paris-XI, France}

\begin{abstract}
We explore possible CP violating effects, coming from the Dirac phase of the Maki-Nakagawa-Sakata-Pontecorvo matrix, on the neutrino degeneracy parameters, at the epoch of Big-Bang nucleosynthesis. 
We first demonstrate the conditions under which such effects can arise. In particular it requires that the initial muon and tau neutrino degeneracy parameters differ.  
Then we solve numerically the kinetic equations for the three flavour neutrino density matrix
with the goal of quantifying the impact of the Dirac phase on $\xi_{\nu_e}$.  
The calculations include the vacuum term, the coupling to matter, the $\nu\nu$ interaction and the collisions. 
Effects on $\xi_{\nu_e}$ up to almost 1.$\%$ and on $Y_p$ of about 0.1$\%$ are found, depending on the initial conditions.
\end{abstract}

\end{frontmatter}
\section{Introduction}
\noindent
One of the major open questions in modern cosmology is the origin of the matter-antimatter asymmetry in our Universe. The bayon asymmetry is nowadays known to be 
$\eta_B\equiv (n_B-n_{\bar{B}})/n_\gamma =6.14 \times 10^{-10}(1.00 \pm 0.04)$
thanks to the measurement of the CMB anisotropies by WMAP \cite{Spergel:2003cb}. Sphalerons effects in baryogenesis and leptogenesis scenarios \cite{Buchmuller:2005eh} can equilibrate cosmic lepton and baryon asymmetries at the same level. Since the lepton asymmetry is only possible in the neutrino sector because of charge conservation, the observation of a non-zero neutrino degeneracy parameter $\xi$ can furnish important information to our understanding of the matter-antimatter asymmetry in the Universe. 

In analogy with $\eta_B$ related to the baryon asymmetry, the total neutrino asymmetry $L_\nu=L_{\nu_e}+L_{\nu_\mu}+L_{\nu_\tau}$ can be quantified by the neutrino chemical potentials $\mu_{\nu_\alpha}$ ($\alpha \equiv e, \mu, \tau$) or, equivalently, the degeneracy parameters $\xi_{\nu_\alpha} \equiv \mu_{\nu_\alpha}/T_\nu$:
\begin{equation}\label{e:nuasym}
L_{\nu_{\alpha}} = \frac{n_{\nu_{\alpha}}-n_{\overline{\nu}_{\alpha}}}{n_{\gamma}}=\frac{\pi^2}{12 \zeta (3)} \left(\frac{T_{\nu_\alpha}}{T_\gamma} \right)^3 \left(\xi_{\nu_{\alpha}}+\frac{\xi^3_{\nu_{\alpha}}}{\pi^2}\right)
\end{equation}
where $n_{\nu_{\alpha}}$  ($n_{\overline{\nu}_{\alpha}}$) are the neutrinos (anti-neutrinos) occupation numbers and $\zeta (3) \simeq 1.202$.
Non-zero electron, muon and tau neutrino degeneracy parameters influence the abundance of light elements produced in
Big-Bang Nucleosynthesis (BBN) in two aspects. While all flavours influence the expansion rate of the Universe, 
by modifying the 
effective number of degrees of freedom, only $\xi_{\nu{e}}$ impacts the neutron/proton ratio, a key parameter for the $^{4}$He abundance. Indeed $^{4}$He, among all the light elements formed during BBN, is the most sensitive one to the neutrino
degeneracy parameters. Extensive work has been performed to extract information on the relic lepton asymmetries either
from Big-Bang Nucleosynthesis, as in e.g.
\cite{Wagoner:1966pv,Kang:1991xa,Esposito:2000hh,Barger:2003rt,Serpico:2005bc,Simha:2008mt}
or from the cosmic microwave background and large scale anisotropies, like in \cite{Lesgourgues:1999wu}. 

Major advances have been performed in neutrino physics in the last ten years.
The change in neutrino flavour content due to oscillations is at present a well established phenomenon.
This implies that the neutrino flavour basis is related to the mass basis
\begin{equation}
  \label{e:basis}
  \psi_{{\nu}_{\alpha}} = \sum_i U_{\alpha i} \psi_i .
\end{equation}
where the unitary Maki-Nakagawa-Sakata-Pontecorvo (MNSP) matrix can be written as
a product of three matrices  $U=T_{23}T_{13}T_{12}$ 
\begin{equation}
\label{e:U}
U  = \left(\matrix{
     1 & 0 & 0  \cr
     0 &  c_{23}  & s_{23} \cr
     0 & - s_{23} &  c_{23} }\right)
 \left(\matrix{
     c_{13} & 0 &  s_{13} e^{-i\delta}\cr
     0 &  1 & 0 \cr
     - s_{13} e^{i\delta} & 0&  c_{13} }\right)
 \left(\matrix{
     c_{12} & s_{12} &0 \cr
     - s_{12} & c_{12} & 0 \cr
     0 & 0&  1 }\right) ,
\end{equation}
\noindent
with $c_{ij} = cos \theta_{ij}$ ($s_{ij} = sin \theta_{ij}$) and $\theta_{12},\theta_{23}$ and $\theta_{13}$ the three neutrino mixing angles. 
These oscillation parameters have been well determined, except for the third neutrino mixing angle $\theta_{13}$ and a possible Dirac CP violating phase\footnote{Note that Majorana phases can also be present. They can influence the neutrinoless double-beta decay half-lives,  while neutrino oscillations are not affected by such phases. For this reason they will not be considered here.}.
If $\theta_{13}$ is close to the Chooz limit, i.e. sin$^2 2\theta_{13}<0.02$, reactor experiments (Double-Chooz, RENO and Daya-Bay) should soon measure this angle \cite{Huber:2009cw}.
The two squared mass differences\footnote{Although the existence of sterile neutrinos is an attractive possibility, here we consider three active neutrino families, in agreement with the ensemble of experimental data.} have been measured with good precision \cite{Amsler:2008zzb}. 
Since the sign of $\Delta m^2_{23}$ has not been determined yet, two mass hierarchies are possible: inverted ($\Delta m^2_{23}>0$) or normal ($\Delta m^2_{23}<0$).  
This is known as the mass hierarchy problem. 
The absolute neutrino mass scale is also still unknown, since neutrino oscillations are only sensitive to mass squared differences. The KATRIN experiment  will soon reach the sub-eV sensitivity \cite{Osipowicz:2001sq} while important indirect limits on the sum of the neutrino masses are obtained using CMB and LSS data (see e.g. \cite{Lesgourgues:2006nd,Fogli:2006yq,Hannestad:2006zg}). Indeed, so far, only indirect effects of cosmological neutrinos have been observed. Their detection represents one of the major future challenges. An interesting possibility has been proposed recently  in \cite{Cocco:2007za}, namely to exploit the capture on radioactive nuclei. This idea has been further investigated in \cite{Lazauskas:2007da,Blennow:2008fh}.  

The observation of CP violation in the neutrino sector is a key open question. The breaking of the CP symmetry
can arise from the presence of a non-zero Dirac $\delta $ phase that renders the $U$ matrix complex Eq.(\ref{e:U}).  Long-term expensive accelerator complex  (super-beams, beta-beams or neutrino factories)
might be required to tackle this issue \cite{Volpe:2006in}.
Therefore it is important to explore complementary avenues and search for indirect effects. For example, recently we have explored possible 
CP violation effects in core-collapse supernovae \cite{Balantekin:2007es}. We have shown that they can arise e.g. if $\nu_{\mu}$ and $\nu_{\tau}$ experience a different 
refractive index in the medium -- due to loop corrections and/or physics beyond the standard model. 
These results have been extended in presence of the neutrino-neutrino interaction in \cite{Gava:2008rp}.
Note that important developments are currently ongoing in the 
study of neutrino propagation in dense media due to temporally evolving density profiles 
\cite{Schirato:2002tg,Fogli:2004ff,Kneller:2007kg,Dasgupta:2005wn,Gava:2009pj,Galais:2009wi}, and 
the neutrino-neutrino interaction, which introduces collective phenomena (see e.g.
\cite{Samuel:1993uw,Sigl:1992fn,Duan:2005cp,Hannestad:2006nj,Raffelt:2007xt}).  
The importance of the latter contribution has been first pointed out in the early Universe context \cite{Dolgov:2002ab,Abazajian:2002qx}.

Several calculations have been performed of the neutrino degeneracy evolution at the BBN epoch including neutrino oscillations \cite{Bell:1998ds,Dolgov:2002ab,Abazajian:2002qx,Mangano:2005cc,Pastor:2008ti}. In \cite{Dolgov:2002ab} it is shown that neutrino oscillations tend to equilibrate the electron, muon and tau neutrino degeneracies. However how much flavour equilibration really holds is still unclear \cite{Pastor:2008ti}. Only in case of flavour equilibration the constraints on $\xi_{\nu_e}$ coming from the abundance of primordial Helium-4 can be translated to the other flavours. In such a case
the limit of $-0.044 < \xi < 0.070$ for all flavours  \cite{Serpico:2005bc}, if the conservative Olive and Skillman analysis is used, with an uncertainty of 
the order of $5 \%$ on $Y_p$ \cite{Olive:2004kq}. Other detailed analysis of the $^{4}$He fraction exist \cite{Izotov:2007ed,Coc:2003ce}.  If the systematic uncertainties, inherent to the helium abundance measurements, are better understood a precision as low as $10^{-3}$ might be reached. Besides, future studies of gravitational lensing distortions on both the temperature and the CMB polarization might reach sensitivities, close to the BBN ones, on the helium fraction, and even at the level of 5.$~10^{-3}$   \cite{Kaplinghat:2003bh}.

In this paper we explore possible CP violating effects, coming from the Dirac phase, on the neutrino degeneracy parameters, at the time of Big-Bang Nucleosynthesis. First we demonstrate analytically the conditions under which there can be such effects. Then we determine numerically the neutrino degeneracies evolution
including, for the first time, a non-zero Dirac phase.
We solve the equations for the three flavour density matrix taking into account the vacuum oscillations, the coupling to the plasma, the neutrino-neutrino interaction and the collisions, using a damping approximation. 

The manuscript is structured as follows. The theoretical formalism is shown in section 2. Section 3 presents our analytical results that define the conditions to have CP effects on the neutrino degeneracy parameters. Section 4 illustrates the numerical results and the potential modifications introduced by the Dirac phase as well as a discussion of the possible implications on the $^{4}$He fraction. Section 5 is the conclusion.

\section{The neutrino evolution equations}
\noindent
The neutrino evolution including oscillations can be determined by using the density matrix:
\be
\label{cosmo1}
\rho_{{\nu}}(p,t)\equiv\left(\begin{array}{ccc}
\rho_{{\nu}_{ee}}  &   \rho_{{\nu}_{e\mu}}    & \rho_{{\nu}_{e\tau}}\     \\
\rho_{{\nu}_{\mu e}}      & \rho_{{\nu}_{\mu\mu}}  & \rho_{{\nu}_{\mu\tau}}     \\
\rho_{{\nu}_{\tau e}}      &   \rho_{{\nu}_{\tau\mu}}    & \rho_{{\nu}_{\tau\tau}}
\end{array} \right)
\ee
in three flavours. Each neutrino state is characterized by the momentum $p$ and the time $t$.
The transposed of $\rho_{{\nu}}(p,t)$, $\bar{\rho}_{{\nu}}(p,t)$, is taken to describe anti-neutrinos.
In an expanding universe, the equations of motion are \cite{Sigl:1992fn}:
\begin{equation}
i(\partial_t -Hp\partial_p)\rho_p= \left[ H_{tot},\rho_p \right] +C(\rho_p) ,
\label{CosmoEOM}
\end{equation}
where the explicit dependence on $t$ is not shown for simplicity, the subscript $p$ refers to the momentum dependence, and
$H_{tot}$ is the total Hamiltonian describing neutrino propagation in the medium.
As long as the expansion rate of the Universe is smaller than the collision rate among the relativistic species, collisions play an important role and drive the system towards equilibrium.
Such contributions, proportional to $G_F^2$ are included here through the collision term $C(\rho)$. 
The cosmic expansion is taken into account through the $Hp\partial_p$ contribution, with $H=\dot{a}(t)/a(t)$.
$a(t)$ is the scale factor that is normalized such as $a \approx 1/T$ at high temperatures or early times. The Hubble constant $H$ is determined through the Friedmann equation $H = \sqrt{8 \pi G\rho/3}$ with $G$ being the gravitational
constant and $\rho$ the total energy density of the relativistic particles. 
By using co-moving variables $x \equiv ma,~y \equiv pa$ ($m$ is an arbitrary mass scale
that we take equal to 1 MeV) Eq.(\ref{CosmoEOM}) becomes adimensional:
\be 
iHx \partial_x \rho_y = \left[H_{tot},\rho_y\right] +C(\rho_y) \label{Cosmo2EOM}
\ee

The
total Hamiltonian describing neutrino propagation involves three contributions
\begin{equation}\label{Htot}
H_{tot} = H_{vac} + H_{mat} + H_{\nu \nu} =  \frac{U M^2 U^{\dagger}}{2p}-\frac{8\sqrt{2}G_F p}{3 m_{W}^2}E+\sqrt2 G_F (\rho-\overline{\rho}),\
\end{equation}
where $G_F$ is the Fermi constant and $m_W$ the W boson mass. 
Anti-neutrino evolution is described by the same Eq.(\ref{Htot}) for $\bar{\rho}_{{\nu}}(p,t)$, but with a minus sign for $H_{vac}$.
The first term is the vacuum oscillation contribution with $M^2=diag(m_1^2,m_2^2,m_3^2)$ and $U$ the MNSP matrix Eq.(\ref{e:U}). 
The second contribution 
is proportional to the energy densities $E$ of charged leptons (electrons, positrons and muons) in the plasma and corresponds to the refractive effects of the medium that neutrinos experience \cite{Sigl:1992fn}. 
Note that the background potential arising due to asymmetries in charged leptons 
is negligible in comparison with the other terms in the Hamiltonian \cite{Abazajian:2002qx}.
The  $\sqrt2 G_F (\rho-\overline{\rho})$ contribution represents the neutrino-neutrino interactions and is meant to be integrated over the neutrino momenta. This non-linear term is responsible for synchronizing the neutrino ensemble, as discussed in \cite{Bell:1998ds,Dolgov:2002ab,Abazajian:2002qx}, 
similarly to what occurs in core-collapse supernovae \cite{Samuel:1993uw,Duan:2005cp,Gava:2008rp}.
Concerning the neutrino scattering with $e^{\pm}$, $\mu^{\pm}$ or among themselves, in principle one should consider the exact collision integral $I_{\nu_{\alpha}}$ 
\cite{Dolgov:1997mb,Pastor:2008ti}  including all relevant two-body weak reactions 
of the type $\nu_{\alpha}(1) +2 \longrightarrow 3 + 4$. Here we
follow \cite{Dolgov:2002ab,Mangano:2005cc,Pastor:2008ti} and use a damping prescription 
of the form\footnote{Note that in \cite{Mangano:2005cc,Pastor:2008ti} the damping prescription is used only for the off-diagonal contributions.} 
\ba\label{e:coll}
C(\rho_{y,\alpha \beta}) & = & -D_{\alpha \beta} \rho_{y,\alpha \beta}\\ \nonumber
C(\rho_{y,\alpha \alpha}) & = & D_{\alpha \alpha}(f(y,\xi_{\alpha})-\rho_{y,\alpha \alpha})
\ea
with $\alpha,\beta =e, \mu, \tau$ and $\xi_{\alpha}$ being the equilibrium solution. 
The coefficients are fixed at
$D_{\alpha \beta}=2(4 sin^4 \theta_W - 2sin^2\theta_W + 2)F_0$ ($\theta_W$ being the Weinberg angle) for 
$\alpha=e$ and $\beta=e,\mu$ or $\tau$, while for all other cases we take
$D_{\alpha \beta}$ or $D_{\alpha\alpha}=2 (2 sin^4 \theta_W + 6)F_0$ with 
$F_0$ as in \cite{Dolgov:2002ab}.

We consider the plasma to be in thermal\footnote{Corrections to the Fermi-Dirac distributions 
have been calculated to be very small \cite{Dolgov:1997mb,Hannestad:1999fj,Mangano:2005cc}.} but not chemical equilibrium.
Therefore, before making the density matrix evolve, we consider  
the neutrino occupation numbers given by Fermi-Dirac distributions $f(y,\xi_{\nu_i})$, characterized 
by the temperature $T$ and the
chemical potentials $\xi$\footnote{Opposite 
chemical potentials are considered for $\nu$ and $\bar{\nu}$.}:
\be
\label{initialcond}
\rho_{{\nu}}(y,t=0)\equiv\left(\begin{array}{ccc}
f(y,\xi_{\nu_e})  &   0    & 0     \\
0      & f(y,\xi_{\nu_{\mu}})  & 0     \\
0      &   0    & f(y,\xi_{\nu_{\tau}})
\end{array} \right).
\ee
 
Before showing the possible impact of  the CP phase $\delta$ on $\xi$, we now try to get an analytical insight on the possible sources for these effects.

\section{Conditions for CP effects on $\xi$ : Analytical results}
\noindent
Let us now demonstrate under which conditions there can be CP violation effects coming from
the Dirac phase at the BBN epoch. To this end, we follow partly the procedure established in Refs.\cite{Balantekin:2007es} and \cite{Gava:2008rp} within the context of core-collapse supernovae. 

For our purpose  it is convenient to work in the $T_{23}$ basis, as shown in Ref \cite{Balantekin:2007es}, but applied to the density matrix Eq.(\ref{cosmo1})
\begin{equation}
\label{e:equatt23}
\tilde{\rho}_{{\nu,y}} = T_{23}^{\dagger} \rho_{{\nu,y}} T_{23} 
\end{equation} 
We also define the useful quantity $\tilde{\rho}_{y,S} = S^{\dagger}\tilde{\rho}_y S$ where the $\delta$ dependence is contained in the unitary diagonal matrix
\begin{equation}
\label{e:S}
S = \left(\matrix{
     1 & 0 & 0  \cr
     0 &  1  & 0 \cr
     0 & 0 &  e^{i\delta} }\right)
\end{equation}
\noindent
Since the MNSP matrix Eq.(\ref{e:U}) can be written as $U=T_{23}ST_{13}^0T_{12}$, we multiply Eq.(\ref{CosmoEOM}) by $T_{23}^{\dagger}$ ($T_{23}$) on the left (right) and get
\be 
iHx \partial_x \tilde{\rho}_{y,S} = \left[ \tilde{H}_{tot},\tilde{\rho}_{y,S} \right]  +C(\tilde{\rho}_{y,S}) \label{e:HtotT23}
\ee
with 
\be\label{e:Htot2T23}
\tilde{H}_{tot} = \frac{T^0_{13}T_{12} M^2 T^{\dagger}_{12}{T^0}^{\dagger}_{13}}{2y}-\frac{8\sqrt{2}G_F y}{3 m_{W}^2}S\tilde{E}S^{\dagger} \\ \newline
 + \sqrt2 G_F (\tilde{\rho}_{y,S} -\tilde{\overline{{\rho}}}_{y,S})
\ee
where $\tilde{E}=T_{23}^{\dagger}~E~T_{23}$ and $E= diag(E_{ee},E_{\mu \mu},0)$. The quantities $E_{ee}$ and $E_{\mu\mu}$ are the energy densities associated with the electrons, positrons and $\mu^+,\mu^-$ respectively.

We now show the conditions under which : i) the initial conditions for  $\tilde{\rho}_S$ and $\tilde{\rho}$ are the same, ii) the evolution equations for  $\tilde{\rho}_S$ Eq.(\ref{e:HtotT23}) are the same as for $\tilde{\rho}$ Eq.(\ref{Cosmo2EOM}).  If both i) and ii) are fullfilled then 
using time discretization by mathematical induction one can show $\tilde{\rho}_S$ is equal to $\tilde{\rho}$ at all times, and therefore the density matrix does not depend on $\delta$ at any time. On the contrary, there can be CP violating effects on $\delta$ and on $\xi$.

Let us consider the initial conditions for $\tilde{\rho}_S$ and for each of the quantities on the r.h.s. of Eq.(\ref{e:HtotT23}) to identify the conditions under which  $\tilde{\rho}_S=\tilde{\rho}(t=0)$, namely when condition i) is satisfied.
At $t=0$ the density matrix is
\be\label{e:mattini}
\tilde{\rho}_{y,S} =\left(\begin{array}{ccc}
f(y,\xi_{\nu_e})  &   0    & 0     \\
0      & c^2_{23}f(y,\xi_{\nu_{\mu}}) + s^2_{23}f(y,\xi_{\nu_{\tau}})  & c_{23}s_{23}(f(y,\xi_{\tau}) -f(y,\xi_{\nu_{\mu}}) )e^{i\delta}     \\
0      &   c_{23}s_{23}(f(y,\xi_{\nu_{\tau}}) -f(y,\xi_{\nu_{\mu}}))e^{-i\delta}    & c^2_{23}f(y,\xi_{\nu_{\tau}})+s^2_{23}f(y,\xi_{\nu_{\mu}})
\end{array} \right)
\ee
This implies that, if the initial muon and tau neutrino degeneracy parameters differ, this engenders 
a dependence on $\delta$ of the density matrix.
Concerning the different terms on the r.h.s. of Eq.(\ref{e:HtotT23}), 
it is first shown in Ref.\cite{Balantekin:2007es} that the vacuum contribution to the Hamiltonian satisfies the factorization $\tilde{H}_{vac}(\delta=0)=S^{\dagger}\tilde{H}_{vac}(\delta)S$ and therefore has no $\delta$ dependence at any time.
The matter related term in Eq.(\ref{e:HtotT23}) is given by:
\be\label{e:mattini}
S\tilde{E}S^{\dagger}(t=0) =\left(\begin{array}{ccc}
E_{ee}  &   0    & 0     \\
0      & -s^2_{23}E_{\mu \mu}  & c_{23}s_{23}E_{\mu \mu}e^{i\delta}     \\
0      &   c_{23}s_{23}E_{\mu \mu}e^{-i\delta}    & -c^2_{23}E_{\mu \mu}
\end{array} \right)
\ee
One can see that, if the presence of muons and anti-muons in the relativistic plasma is not neglected at this epoch of the Universe evolution, then this will introduce a source of CP-violation. 
Note that it has been explicitly shown in Ref.\cite{Gava:2009gt} that a difference in the 
$\nu_{\mu}$ and $\nu_{\tau}$ refractive indeces induces a dependence on $\delta$ in the $\nu_e$ channel. 
However, the more the temperature goes down, the less the $E_{\mu \mu}$ term will be important. 
Concerning the $\nu\nu$ interaction contribution at initial time, one has before integrating over the neutrino momenta :
\be\label{e:nunu}
\tilde{\rho}_{S} -\tilde{{\bar{\rho}}}_{S} =
n_\gamma \left(\begin{array}{ccc}
L_{\nu_e}  &   0    & 0     \\
0      & c^2_{23}L_{\nu_{\mu}}+s^2_{23}L_{\nu_{\tau}} & c_{23}s_{23}(L_{\nu_{\mu}}-L_{\nu_{\tau}})e^{i\delta}     \\
0      &   c_{23}s_{23}(L_{\nu_{\mu}}-L_{\nu_{\tau}})e^{-i\delta}    & s^2_{23}L_{\nu_{\mu}}+c^2_{23}L_{\nu_{\tau}}
\end{array} \right)
\ee
where $L_{\nu_i}$ is the $i$ flavour lepton asymmetry and $n_\gamma$ the photon number density.
Any lepton flavour asymmetry between muon and tau neutrinos introduces a CP dependence of the neutrino-neutrino interaction Hamiltonian. 
Finally, one has that  the collision term $C(\rho(p,t))$ has no $\delta$ dependence since
$\rho_{\nu_i \nu_i}=f(y,\xi_i)$ and the $\rho_{\nu_i \nu_j}=0$ for $i \neq j$ making such a term equal to zero in any basis initially. 

Let us now show that condition ii) holds. As shown in \cite{Balantekin:2007es} the matter term is equal in both equations if and only if one can neglect the $E_{\mu \mu}$ contribution. Concerning the neutrino-neutrino interaction term, since the corresponding Hamiltonian has a linear dependence in the density matrix 
$S^{\dagger}\tilde{H}_{\nu\nu}(\delta)S=\tilde{H}_{\nu\nu}(\delta=0)$ at all times, as demonstrated by mathematical induction in Ref.\cite{Gava:2008rp}, using the Liouville-Von Neumann equation, in the supernova context. Here this requires the muon and tau neutrino asymmetries being equal ($L_{\nu_{\mu}}=L_{\nu_{\tau}}$) at initial time. Finally the collision term, here treated in the damping approximation, is also a linear function of the density matrix, therefore its dependence is the same if $\tilde{\rho}_S$ and $\tilde{\rho}$ are used. 

In conclusion, if $E_{\mu \mu}$ is negligeable and $L_{\nu_{\mu}}=L_{\nu_{\tau}}$ at $t=0$, then
the Hamiltonian that governs the evolution of the density matrix with and without a dependence of the Dirac phase is the same. As a consequence, the evolution of   $\tilde{\rho}_S$ and $\tilde{\rho}$ is the same. This implies that, under such conditions,  
$\rho_{\nu_e\nu_e}(\delta)=\rho_{\nu_e\nu_e}(\delta=0)$, and therefore 
$\xi_{\nu_e}(\delta)=\xi_{\nu_e}(\delta=0)$ at all times.

Summarizing, we have demonstrated that possible CP violating effects can arise from the Dirac phase $\delta$
if, initially, there is a difference between the muon and tau neutrino occupation numbers and/or degeneracy parameters. If the (anti)muon contribution to the energy density is non-negligible this can also, in principle, engender
CP violating effects.

\section{Impact of the CP phase on $\xi_{\nu_e}$ : Numerical results}
\noindent
The goal here is to quantify possible CP effects on the neutrino degeneracies
just before Big-Bang nucleosynthesis, having in mind that this can e.g. potentially impact the Helium-4 fraction. 
The numerical results we present are obtained by solving 
Eqs.(\ref{Cosmo2EOM}-\ref{e:coll}) for the density matrix
and the initial conditions Eq.(\ref{initialcond}).
The neutrino degeneracies are then obtained using Eq.(\ref{e:nuasym}). 
We will give the variations on $\xi_{\nu_e}$ induced by $\delta$ since this is the relevant quantity to quantify the effect on the helium-4 fraction.

In our calculations, the oscillation
parameters are fixed at the values 
$\Delta m^2_{12}= 8 \times 10^{-5}$eV$^2$, sin$^2 2\theta_{12}=0.83$ and
$\Delta m^2_{23}= 3 \times 10^{-3}$eV$^2$, sin$^2 2\theta_{23}=1$ for
the solar and atmospheric differences of the squared mass differences and
mixings, respectively \cite{Amsler:2008zzb}. For the third still unknown neutrino mixing angle
$\theta_{13}$, we have taken either a large value close to the Chooz limit, namely 
sin$^2 2\theta_{13}=0.19$ at 90 $\%$ C.L., or a small value of 
sin$^2 2\theta_{13}=3 \times 10^{-4}$.

\begin{figure}[t]
\centerline{\includegraphics[scale=0.3,angle=0]{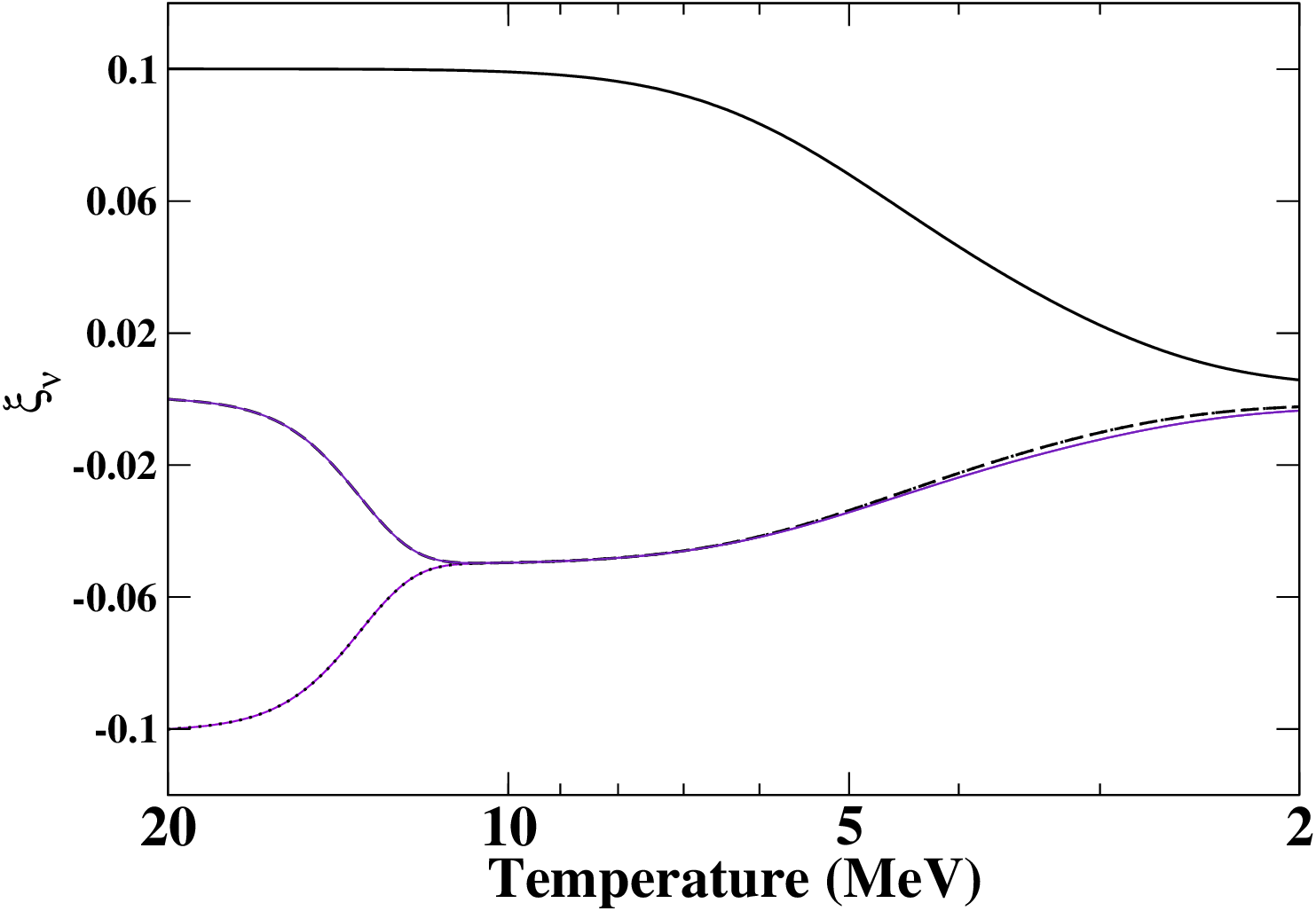}}
\vspace{.25cm}
\caption{Neutrino degeneracy parameters as a function of the temperature for $\delta=0^{\circ}$ and $\delta=180^{\circ}$ , at the BBN epoch. The initial conditions are set
at $\xi_{\nu_e}=0.1$, $\xi_{\nu_{\tau}}=-0.1$ and $\xi_{\nu_{\mu}}=0$. Results obtained solving Eqs. (\ref{Cosmo2EOM}-\ref{e:coll})
and the initial conditions Eq.(\ref{initialcond}) but
without including the neutrino-neutrino interaction and the diagonal collision terms. The third neutrino mixing angle is taken at the Chooz limit. Here the curves corresponding to a non-zero Dirac phase are indistinguishable from those with a zero value.}
\label{fig1}
\end{figure}

\begin{figure}[t]
\centerline{\includegraphics[scale=0.4,angle=0]{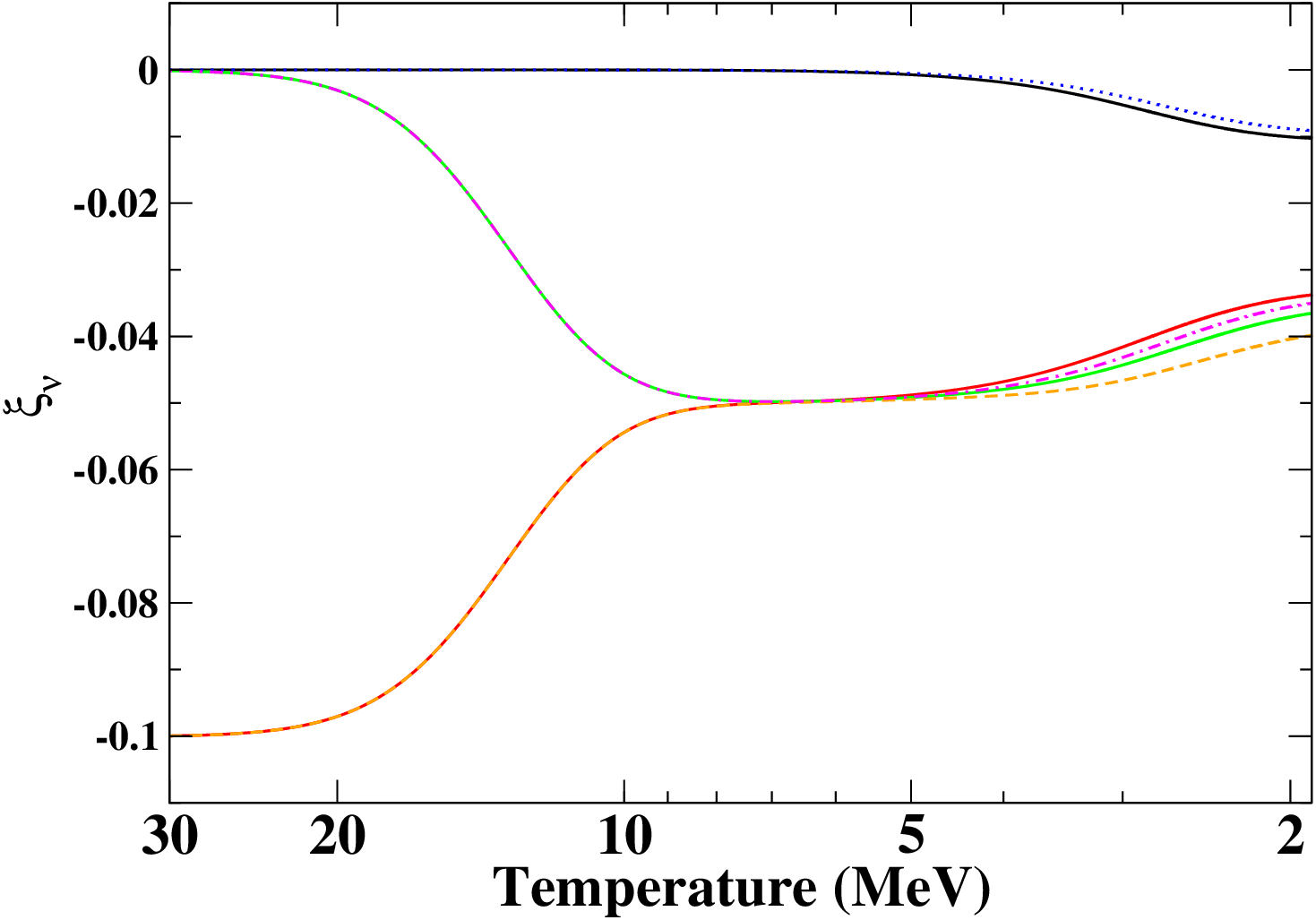}}
\vspace{.25cm}
\caption{CP effects on the $\xi$ as a function of the temperature.  The initial conditions here are 
$\xi_{\nu_e}=\xi_{\nu_{\tau}}=0.$ and $\xi_{\nu_{\mu}}=-0.1$. 
The calculations include collision terms but do not include the neutrino-neutrino interaction. The mixing angle $\theta_{13}$ is taken at the Chooz limit.}
\label{fig2}
\end{figure}

\begin{figure}[t]
\centerline{\includegraphics[scale=0.3,angle=0]{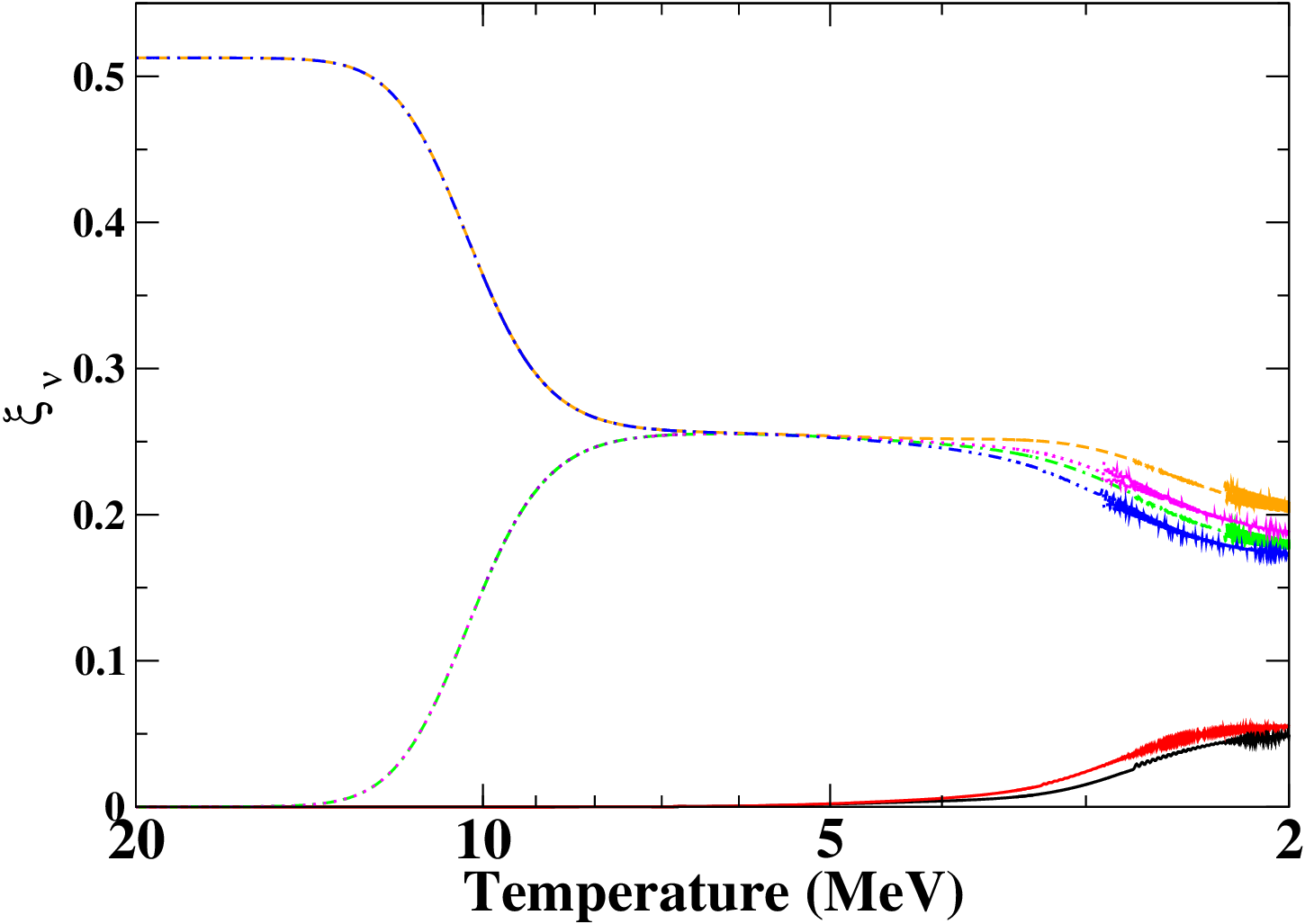}}
\vspace{.25cm}
\caption{Neutrino degeneracy parameters, as a function of the temperature, when the initial conditions are taken equal to $\xi_{\nu_e}=\xi_{\nu_{\mu}}=0$  and $\xi_{\nu_{\tau}}=0.5$.
The results correspond to $\xi_{\nu_e}$ for $\delta=180^{\circ}$ and $\delta=0^{\circ}$ (lower lines), to $\xi_{\nu_{\tau}}$ for $\delta=180^{\circ}$ (dashed) and $\delta=0^{\circ}$ (dot-dot-dashed) and to $\xi_{\nu_{\mu}}$ for $\delta=180^{\circ}$ (dotted) and $\delta=0^{\circ}$ (dot-dashed). The calculations include the vacuum oscillation and matter term, the neutrino-neutrino interaction and the collisions. }
\label{fig3}
\end{figure}

\begin{figure}[t]
\centerline{\includegraphics[scale=0.3,angle=0]{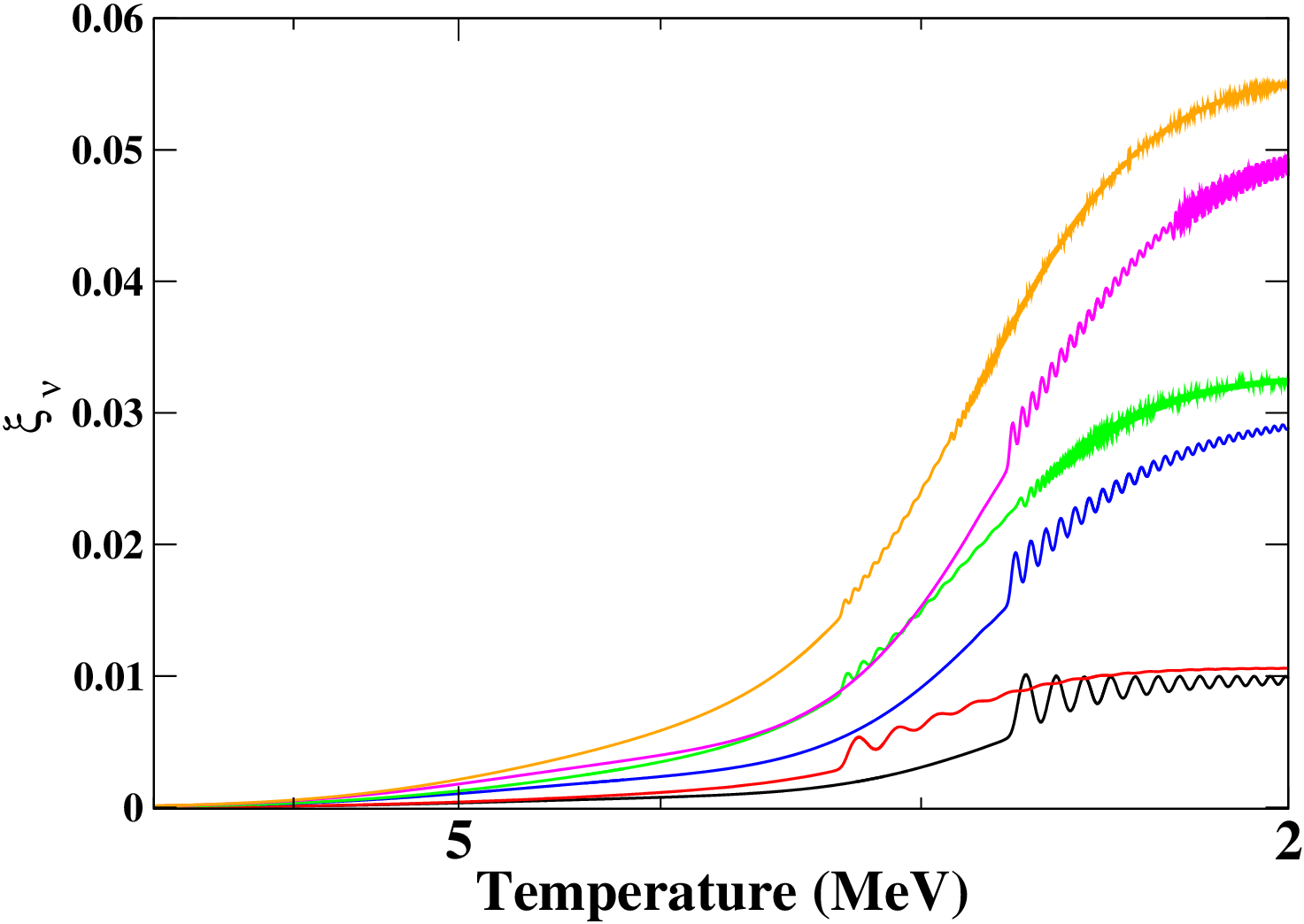}}
\vspace{.25cm}
\caption{Neutrino degeneracy parameters $\xi_{\nu_e}$, as a function of the temperature, 
for different values of the initial degeneracy parameters. The lines correspond to $\delta=180^{\circ}$ in comparison with the case $\delta=0^{\circ}$, for the initial values of $\xi_{\nu_{\tau}}=0.1$ (lower), $0.3$ (middle) and $0.5$ (upper curves). The values of $\xi_{\nu_e}$ and $\xi_{\nu_{\mu}}$  at $t=0$ are set to zero, for all cases.}
\label{fig4}
\end{figure}

Figures \ref{fig1}-\ref{fig4} present the evolution of the neutrino degeneracy parameters as a function of the temperature. Figure \ref{fig1} shows the results obtained for $\xi$ with $\delta=0^{\circ}$ and $\delta=180^{\circ}$, without the inclusion of the neutrino-neutrino contribution to the total Hamiltonian Eq.(\ref{e:Htot2T23}). Only the off-diagonal contributions to the collision term are included.
One can see that the effect of $\delta$ are very small, the curves for a non-zero $\delta$ value being indistiguishable from those for $\delta=0^{\circ}$, as we have been verifying for different initial conditions.
For example, if at initial time $\xi_{\nu_e}=-0.3$, $\xi_{\nu_{\mu}}=0.3$ and $\xi_{\nu_{\tau}}=0$, the variation induced by the phase is  
$\Delta \xi_{\nu_e} = \xi_{\nu_e}(\delta) - \xi_{\nu_e}(\delta=0)=4 \times 10^{-6}$, while for  $\xi_{\nu_e}=-0.5$, $\xi_{\nu_{\mu}}=0.5$  and $\xi_{\nu_{\tau}}=0$, we obtain $\Delta \xi_{\nu_e} =10^{-5}$.

Figures \ref{fig2} and \ref{fig3}-\ref{fig4} show the CP effect on $\xi$ without and with the neutrino-neutrino contribution\footnote{The calculations including the neutrino-neutrino contribution have been performed following the procedure used in \cite{Dolgov:2002ab,Pastor}.}, respectively. Here both off-diagonal and diagonal collision terms are included\footnote{We use the approximation that the equilibrium $\xi$ are kept at the initial value \cite{Dolgov:2002ab}.}.
Results are shown with $\delta=0^{\circ}$ and $\delta=180^{\circ}$. 
In Figure \ref{fig2}
the effect of the CP phase on $\xi_{\nu_e}$ is at the level of $10^{-3}$.
The degeneracy parameters are compatible with the bounds valid if total flavour equilibration is assumed. In Fig.\ref{fig3} the effect is $\Delta \xi_{\nu_e} = 6 \times 10^{-3}$. Note that, for such initial conditions, while
the value of $\xi_{\nu_e}$ at decoupling is compatible with the present bound from BBN; 
the other degeneracy parameters are compatible with the bounds valid without assuming total flavour equilibration.  
Figure \ref{fig4} presents $\Delta \xi_{\nu_e}$ for different initial conditions, showing that the variations induced by $\delta$ increase, if $\xi_{\nu_e}$ at the temperature of neutrino decoupling is larger.

In all the calculations performed we have found that the modifications produced by
the CP effects increase for a maximal phase and if the third neutrino mixing angle is large. Note that the CP effects arising from the presence of a non-zero $E_{\mu\mu}$
have been found to be completely negligeable. 
The inclusion of the non-linear $\nu\nu$ contribution, in general, reduces the phase impact. This can be qualitatively understood. In fact such term synchronizes the neutrino ensemble and freezes neutrino flavour conversion. A similar reduction of CP effects on the neutrino oscillation probabilities has been found in the context of core-collapse supernovae, when the neutrino coupling to neutrinos is included (see Figure 3 of Ref.\cite{Gava:2008rp}). 
Finally, our results show that the effect 
depends on the diagonal collision terms. Clearly, a definite conclusion on the quantitative effects needs the inclusion of the exact collision integrals. This will be the object of future investigations.

As far as the impact on the on $^{4}$He fraction is concerned,
an estimate of the CP effects can be obtained using the simple relation $\Delta Y_p=-0.2 \Delta \xi_{\nu_e}$ \cite{Kneller:2004jz}. In fact, a modification of the order of $\Delta \xi_e$ of a several $10^{-3}$, as we have found for some initial conditions, modifies $Y_p$ at most by about $10^{-3}$.  This is within the uncertainty from BBN observations, although  the present uncertainty on the $^{4}$He fraction might be narrowed down at the level of $Y_p < 0.005 $ in the future \cite{Kaplinghat:2003bh}.
  
\section{Conclusions}
\noindent
We have explored the impact of the Dirac CP phase of the MNSP matrix on the neutrino degeneracy
parameters at the Big-Bang nucleosynthesis epoch. First we have established analytically the conditions under which there can be possible CP effects coming from this phase and shown, in particular, that these are present if there is a difference between the initial muon and tau neutrino degeneracy parameters. To quantify such
effects we have numerically solved the evolution equation for the density matrix in three flavours, including mixings in
vacuum, coupling to matter, the $\nu\nu$ interaction and collisions (in the damping approximation). 
We have found that, depending on the initial conditions
for the neutrino degeneracy parameters, which are an unknown,
modifications up to almost $1.\%$ and $0.1\%$ might be present on $\xi_{\nu_e}$ and $Y_p $ respectively, when the $\nu\nu$ interaction and the collision terms are included. 

\vspace{.4cm}

\noindent
{\bf Acknowledgement}
\vspace{.1cm}

\noindent
We thank Sergio Pastor and Georg Raffelt for providing us with important information,  Alain Coc, James Kneller and Julien Serreau for useful discussions.

\end{document}